\documentstyle[12pt]{article}
\makeatletter

\@addtoreset{equation}{section}
\makeatother

\newcommand{\beq}{\begin{equation}}
\newcommand{\eeq}{\end{equation}}
\newcommand{\beqa}{\begin{eqnarray}}
\newcommand{\eeqa}{\end{eqnarray}}

\title{Cancellation of UV Divergences in the ${\cal N}=4$ SUSY Nonlinear Sigma Model in Three Dimensions}
\author{Takeo Inami,
Yorinori Saito
and
Masayoshi Yamamoto\\
\it Department of Physics,
Faculty of Science and Engineering\\
\it Chuo University,
1-13-27 Kasuga, Bunkyo-ku, Tokyo 112-8551, Japan}
\date{\,}
\begin{document}
\maketitle

\begin{center}
\begin{abstract}We study the UV properties of the three-dimensional ${\cal N}=4$ SUSY nonlinear sigma model whose target space is $T^*(CP^{N-1})$ (the cotangent bundle of $CP^{N-1}$) to higher orders in the $1/N$ expansion. We calculate the $\beta$-function to next-to-leading order and verify that it has no quantum corrections at leading and next-to-leading orders.
\end{abstract}
\end{center}

\newpage

\section{Introduction}
Three-dimensional nonlinear sigma models have special properties regarding UV divergences. They are non-renormalizable theories in the sense of perturbation expansion, but they are renormalizable in the $1/N$ expansion \cite{Arefeva, Vasilev}. The three-dimensional $O(N)$ and $CP^{N-1}$ nonlinear sigma models were studied to next-to-leading order in $1/N$ and their $\beta$-functions were determined to this order \cite{Rosenstein, Gracey, Cant}. 

An important feature of SUSY field theories is weaker quantum corrections, particularly UV divergences in the perturbation expansion. We want to pose the question: ``Do SUSY field theories have this feature in the $1/N$ expansion? Does the model with higher ${\cal N}$ extended SUSY have better UV property?''  We address ourselves to these questions in extended SUSY nonlinear sigma models in three dimensions.

Some works have been made in this direction. In the ${\cal N}=1$ SUSY $O(N)$ nonlinear sigma model in three dimensions, the next-to-leading order term in the $\beta$-function turned out to be absent modulo power divergences in the $1/N$ expansion \cite{Koures} and in the critical exponent technique \cite{Gracey2}. In the ${\cal N}=2$ SUSY $CP^{N-1}$ model in three dimensions, the next-to-leading order term in the $\beta$-function was found to vanish \cite{Inami, Gracey3}. In the ${\cal N}=4$ SUSY nonlinear sigma model in three dimensions whose target space is $T^*(CP^{N-1})$ (the cotangent bundle of $CP^{N-1}$), the $\beta$-function was found to receive no quantum corrections at leading order \cite{Inami2}. Curiously, these results in low orders of $1/N$ are reminiscent of the UV properties in perturbation of ${\cal N}=1,2$ and $4$ SUSY gauge theories in four dimensions.

We have initiated a study of the UV properties of the ${\cal N}=4$ SUSY $T^*(CP^{N-1})$ model in three dimensions in higher orders of the $1/N$ expansion. Nonlinear sigma models in three dimensions are plagued by a number of power divergences in the cutoff $\Lambda$. We investigate how such UV divergences may combine to cancel out in the model. To this end we use the cutoff regularization. In this letter we present the result of the computation of the $\beta$-function to next-to-leading order in $1/N$. We have previously shown that the $\beta$-function at leading order receives no quantum corrections in the saddle point evaluation \cite{Inami2}. We examine whether this remarkable property will persist at higher orders.

\section{The Model}

We consider the ${\cal N}=4$ SUSY $T^*(CP^{N-1})$ model in three dimensions \cite{Inami2}. The model can be constructed from the ${\cal N}=2$ model in four dimensions \cite{Curtright} by dimensional reduction. We follow \cite{Curtright} and use the component language. The model consists of $2N$ complex scalar fields $\phi^\alpha_i (x)$ ($i=1,2$; $\alpha=1,\dots,N$), $2N$ Dirac fields $\psi^\alpha_i (x)$ (the superpartners of $\phi^\alpha_i$) and  auxiliary fields $\sigma (x)$ (real scalar), $\tau (x)$ (complex scalar), $A_\mu (x)$ ($U(1)$ vector). The Lagrangian is given by \cite{Inami2}
\beqa
{\cal L}_1&=&\overline{D_\mu\phi^\alpha_i}D_\mu\phi^\alpha_i
+i\bar{\psi}^\alpha_i\gamma_\mu D_\mu\psi^\alpha_i
-\tau\bar{\psi}^\alpha_1\psi^\alpha_2-\bar{\tau}\bar{\psi}^\alpha_2\psi^\alpha_1
\nonumber\\
&&+\sigma(\bar{\psi}^\alpha_1\psi^\alpha_1-\bar{\psi}^\alpha_2 \psi^\alpha_2)+(\bar{\tau}\tau+\sigma^2)\bar{\phi}^\alpha_i\phi^\alpha_i,
\label{3daction}
\eeqa
with the constraints
\beqa
&&\bar{\phi}^\alpha_1 \phi^\alpha_1-\bar{\phi}^\alpha_2\phi^\alpha_2=N/g,
~~~\bar{\phi}^\alpha_1\phi^\alpha_2=0,
\label{3dconstraint1}\\
&&\bar{\phi}^\alpha_1\psi^\alpha_1-i\phi^\alpha_2\psi^{\alpha *}_2=0,
~~~\bar{\phi}^\alpha_1\psi^\alpha_2+i\phi^\alpha_2\psi^{\alpha *}_1=0.
\label{3dconstraint2}
\eeqa
We use the Euclidean metric and $D_\mu=\partial_\mu+iA_\mu$ ($\mu=1,2,3$). The symbol $\gamma^\mu$ is the Dirac matrices in three dimensions. They are given by  $\gamma^1=i\sigma_2$, $\gamma^2=i\sigma_3$ and $\gamma^3=i\sigma_1$. Simple dimensional reduction assures that the model (\ref{3daction}) inherits  ${\cal N}=4$ SUSY from the four-dimensional ${\cal N}=2$ SUSY model \cite{Curtright}.

The constraints (\ref{3dconstraint1}) and (\ref{3dconstraint2}) may be expressed as $\delta$-functionals. This introduces a real scalar $\alpha (x)$, a complex scalar $\beta (x)$ and two complex spinors $c (x)$ and $e (x)$ as the Lagrange multiplier fields:
\beqa
{\cal L}&=&{\cal L}_1-\alpha( \bar{\phi}^\alpha_{1}\phi^\alpha_{1}-\bar{\phi}^\alpha_{2}\phi^\alpha_{2}-N/g )\nonumber\\
&&-\beta\bar{\phi}^\alpha_{1}\phi^\alpha_{2}-\bar{\beta}\bar{\phi}^\alpha_{2}\phi^\alpha_{1}
\nonumber\\
&&+\bar{\phi}^\alpha_{1}\bar{c}\psi^\alpha_{1}+\phi^\alpha_{1}\bar{\psi}^\alpha_{1} c+i\bar{\phi}^\alpha_{2}\bar{c}^* \psi^\alpha_{2}-i\phi^\alpha_{2}\bar{\psi}^\alpha_{2} c^*
\nonumber\\
&&+\bar{\phi}^\alpha_{1}\bar{e}\psi^\alpha_{2}+\phi^\alpha_{1}\bar{\psi}^\alpha_{2} e-i\bar{\phi}^\alpha_{2}\bar{e}^* \psi^\alpha_{1}
+i\phi^\alpha_{2}\bar{\psi}^\alpha_{1} e^*.
\label{eq:lag0}
\eeqa
The sets of the fields $(A_\mu, c, \sigma, \alpha)$ and $(\tau, e, \beta)$ are the components of the ${\cal N}=2$  $U(1)$ vector multiplet in the Wess-Zumino gauge and the ${\cal N}=2$  Lagrange multiplier multiplet respectively, which are obtained by dimensional reduction of the four-dimensional ${\cal N}=2$ model in the superfield formulation \cite{Rocek}.

The vacuum of the model is determined by the expectation values of the scalar fields $\phi_i^\alpha$. Taking account of the constraints (\ref{3dconstraint1}), we set
\beq
\langle\stackrel{\rightarrow}{\phi_1}\rangle=(0,\cdots,\sqrt{N}r),~~~\langle\stackrel{\rightarrow}{\phi_2}\rangle=(0,\cdots,\sqrt{N}s,0). 
\eeq
The values of $r$ and $s$ are fixed from the saddle point conditions. Because of the constraints (\ref{3dconstraint1}), only the broken $SU(N)$ phase is allowed. The vacuum expectation values $r$ and $s$ are related to the coupling constant as $r^2-s^2=1/g$. We study the UV properties of the model by setting $s=0$. We should obtain the same result regarding the UV property of the model for other values of $r$ and $s$. Performing the shift
\beq
\phi^N_1\to\phi^N_1+\sqrt{N}r,
\label{phishift}
\eeq
in (\ref{eq:lag0}), we obtain the Lagrangian
\beqa
{\cal L}^{\prime}&=&\overline{\phi_{i}^\alpha}( -\partial^2-iA_{\mu} \stackrel{\leftrightarrow}{\partial_\mu}+A^2)\phi_{i}^\alpha+(\bar{\tau}\tau+\sigma^2)\bar{\phi}^\alpha_{i}\phi^\alpha_{i}
\nonumber\\
&&-\beta\bar{\phi}^\alpha_{1}\phi^\alpha_{2}-\bar{\beta}\bar{\phi}^\alpha_{2}\phi^\alpha_{1}-\alpha(\bar{\phi}^\alpha_{1}\phi^\alpha_{1}-\bar{\phi}^\alpha_{2}\phi^\alpha_{2}-N/g)
\nonumber\\
&&+\overline{\psi}_{i}^\alpha\left( i\ooalign{\hfil/\hfil\crcr$\partial$}-\ooalign{\hfil/\hfil\crcr$A$}\right)\psi_{i}^\alpha-\tau\bar{\psi}^\alpha_{1}\psi^\alpha_{2}-\bar{\tau}\bar{\psi}^\alpha_{2}\psi^\alpha_{1}+\sigma(\bar{\psi}^\alpha_{1}\psi^\alpha_{1}-\bar{\psi}^\alpha_{2}\psi^\alpha_{2})
\nonumber\\
&&+\bar{\phi}^\alpha_{1}\bar{c}\psi^\alpha_{1}+\phi^\alpha_{1}\bar{\psi}^\alpha_{1} c+i\bar{\phi}^\alpha_{2}\bar{c}^* \psi^\alpha_{2}-i\phi^\alpha_{2}\bar{\psi}^\alpha_{2} c^*
\nonumber\\
&&+\bar{\phi}^\alpha_{1}\bar{e}\psi^\alpha_{2}+\phi^\alpha_{1}\bar{\psi}^\alpha_{2} e
-i\bar{\phi}^\alpha_{2}\bar{e}^*\psi^\alpha_{1}+i\phi^\alpha_{2}\bar{\psi}^\alpha_{1} e^*\nonumber\\
&&+N r^2 (A^2+\bar{\tau}\tau+\sigma^2-\alpha)
\nonumber\\
&&+\sqrt{N}\bar{r}(-i\partial^\mu A_{\mu}+A^2+\bar{\tau}\tau+\sigma^2-\alpha)\phi^N_{1}\nonumber\\
&&+\sqrt{N}r \bar{\phi}^N_{1}(i\partial^\mu A_{\mu}+A^2+\bar{\tau}\tau+\sigma^2-\alpha)
\nonumber\\
&&+\sqrt{N}(r\beta\phi^N_{2}+\bar{r}\bar{\beta}\bar{\phi}^N_{2}+\bar{r}\bar{c}\psi^N_{1}+r\bar{\psi}^N_{1} c+\bar{r}\bar{e}\psi^N_{2}+r\bar{\psi}^N_{2} e).\label{eq:lag1}
\eeqa
The prescription of computing quantum corrections in the $1/N$ expansion is the same as that in the $CP^{N-1}$ model \cite{Arefeva2}. We need the effective propagators of the auxiliary fields. They are given by \cite{Inami2}
\beqa
&&D^A_{\mu\nu}(p)=\frac{1}{N}\frac{4}{\sqrt{p^2}+8 r^2}
\left(\delta_{\mu\nu}-\frac{p_\mu p_\nu}{p^2}\right),
\nonumber\\
&&D^\sigma(p)=\frac{1}{N}\frac{4}{\sqrt{p^2}+8 r^2},
~~~D^\tau(p)=\frac{1}{N}\frac{8}{\sqrt{p^2}+8 r^2},
\nonumber\\
&&D^\alpha(p)=-\frac{1}{N}\frac{4p^2}{\sqrt{p^2}+8 r^2},
~~~D^\beta(p)=-\frac{1}{N}\frac{8p^2}{\sqrt{p^2}+8 r^2},
\nonumber\\
&&D^c(p)=\frac{1}{N}\frac{8p\!\!\!/}{\sqrt{p^2}+8 r^2},
~~~D^e(p)=\frac{1}{N}\frac{8p\!\!\!/}{\sqrt{p^2}+8 r^2}.
\label{propagator}
\eeqa
We have used the Landau gauge in deriving $D^A_{\mu\nu} (p)$.

\section{The $\beta$-Function}
The bare quantities denoted by the subscript 0 are related to renormalized quantities by 
\beqa
&&\phi_{i,0}=(Z_{\phi i})^{1/2} \phi_{i},
~~~\psi_{i,0}=(Z_{\psi i})^{1/2} \psi_{i},
~~~g_{0}=Z_g g \label{g}\\
&&\varphi_{0}=Z_\varphi \varphi,~~~\varphi=\alpha, \beta, \sigma, \tau, A_\mu, c, e.
\eeqa
We decompose the bare Lagrangian ${\cal L}_0$ into the renormalized part ${\cal L}$ and the counterterm  Lagrangian ${\cal L}_{\rm CT}$, ${\cal L}_0={\cal L}+{\cal L}_{\rm CT}$. ${\cal L}_0$ and ${\cal L}$ are written in terms of the bare and renormalized quantities, respectively. ${\cal L}_0$ is exactly of the same form as ${\cal L}$. ${\cal L}_{\rm CT}$ is designed to eliminate all UV divergences in n-point functions due to loop effects. Because of the shift (\ref{phishift}), it is given by
\beqa
{\cal L}_{\rm CT}^{\prime}&=&-C_1\phi_1\partial^2\phi_1-C_2\alpha(\overline{\phi}_1\phi_1+Nr^2)+C_g \alpha N/g \nonumber \\
&&+C_3\overline{\psi}_1 i\ooalign{\hfil/\hfil\crcr$\partial$}\psi_1+C_4\sigma\overline{\psi}_1\psi_1+\cdots,
\eeqa
where
\beqa
&&C_1=Z_{\phi 1}-1,~~~C_2=Z_{\alpha} Z_{\phi 1}-1,~~~C_g=Z_{\alpha}Z_{g}^{-1}-1,\\
&&C_3=Z_{\psi 1}-1,~~~C_4=Z_{\sigma} Z_{\psi 1}-1,~~~\cdots.\label{zfac}
\eeqa
The $Z$ and $C$ factors are expanded in $1/N$ as $Z=Z^{(0)}+Z^{(1)}+\cdots$ and $C=C^{(0)}+C^{(1)}+\cdots$.

Before discussing the main result of our study, we summarize the result in leading order \cite{Inami2}. There are only a few kinds of loop diagrams in leading order: the tadpole, self-energy and three-point vertex function of the auxiliary fields (without containing $\phi_i$ and $\psi_i$). We are concerned with these diagrams.

$Z_g^{(0)}$ can be obtained by computing the one loop $\alpha$-tadpole contributing to the one-point vertex function ${\mit\Gamma}_{\alpha}$. We find from (\ref{eq:lag1}) that $\phi_1$ and $\phi_2$ loops contribute to this diagram. Since the $\phi_1$ and $\phi_2$ modes contribute with opposite signs, this diagram is zero. This cancellation mechanism of UV divergences is the same as that of the two-dimensional ${\cal N}=4$ SUSY $T^*(CP^{N-1})$ model in the usual perturbation expansion \cite{Curtright}. The three-point vertex function of the auxiliary fields vanish identically as in the $CP^{N-1}$ model \cite{Arefeva2} and the four-fermion model \cite{Rosenstein2} in three dimensions. The same argument holds for the $\sigma$-tadpole \cite{Rosenstein2}.

In fact, the leading-order tadpole diagrams have already been accounted for by the saddle point conditions and so we need not discuss them except to say that these tadpole diagrams should be considered illegal as subdiagrams. Likewise, the self-energy diagrams of the auxiliary fields are also illegal subdiagrams because they are taken into account by the effective propagators (\ref{propagator}) and are all finite. 

Therefore, the model is finite to leading order in $1/N$:
\beq
Z^{(0)}=1,
\eeq
for all factors. In particular, it implies that the $\beta$-function receives no quantum corrections at leading order. However the $\beta$-function receives the trivial tree level contribution; the dimensionless coupling constant ${\tilde g} =\mu g$ (the renormalization scale $\mu$ is of dimension one) depends on $\mu$ at tree level. The $\beta$-function at leading order is therefore given by
\begin{eqnarray}
\beta^{(0)}({\tilde g})={\tilde g}.
\end{eqnarray}

\begin{figure}[t]
\hspace*{20mm}
\vspace*{-14mm}
\input{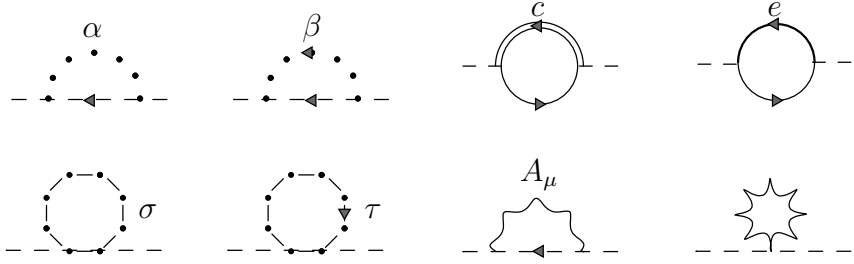}
\caption{Next-to-leading order diagrams contributing to renormalization of $\phi_1$. We denote the propagators of $\phi$ and $\psi$ by dashed and thin solid lines, respectively.}
\label{self}
\end{figure}

We now proceed to next-to-leading order in $1/N$. We have calculated the next-to-leading order corrections to the self-energies of bosons $\phi_i$ and fermions $\psi_i$  and those to the three-point vertex functions ${\mit\Gamma}_{\alpha \bar{\phi}\phi}$ and ${\mit\Gamma}_{\sigma \bar{\psi}\psi}$. Next-to-leading order diagrams contributing to renormalization of $\phi_1$ are shown in Fig.~\ref{self}. These self-energy diagrams contain UV power divergences, but they cancel out in the sum of all diagrams. This is  because the power-divergent terms cancel between the loops of bosons and fermions of the same multiplet due to SUSY. The two loop diagrams contributing to ${\mit\Gamma}_{\alpha \bar{\phi}\phi}$ (${\mit\Gamma}_{\sigma \bar{\psi}\psi}$) also contain UV power divergences. We find from (\ref{eq:lag1}) that $\phi_1$ ($\psi_1$)  and $\phi_2$ ($\psi_2$) loops contribute to these diagrams. Since the $\phi_1$ ($\psi_1$) and $\phi_2$ ($\psi_2$) modes contribute with opposite signs, the each of these diagrams is zero. The remaining logarithmic divergences are removed by the $Z$ factors in next-to-leading order. Therefore we obtain
\beq
Z_{\phi i}^{(1)}=-{2\over N\pi^2}\ln {\Lambda \over \mu},~~~Z_{\psi i}^{(1)}=-{6\over N\pi^2}\ln {\Lambda \over \mu},~~~Z_{\alpha}^{(1)}=Z_{\sigma}^{(1)}=0.\label{zfac2}
\eeq

The boson and fermion wave-function renormalization constants $Z_\phi$ and $Z_\psi$ should be the same in a manifestly SUSY calculation scheme.  $Z_\phi^{(1)}$ and $Z_\psi^{(1)}$ we have obtained turn out to be unequal. We can think of two possible causes for this disagreement. i) We have used the component language taking the Wess-Zumino gauge for the $U(1)$ vector multiplet. Supersymmetry is broken by this choice. The disagreement of the boson and fermion $Z$ factors have been noted in the perturbative calculation in the SUSY Yang-Mills theory in the Wess-Zumino gauge \cite{Jones}. ii) The momentum cut-off regularization is likely to break SUSY due to asymmetric treatment of boson and fermion loop momenta \cite{Inami3}.
\begin{figure}[t]
\hspace*{15mm}
\vspace*{3mm}
\input{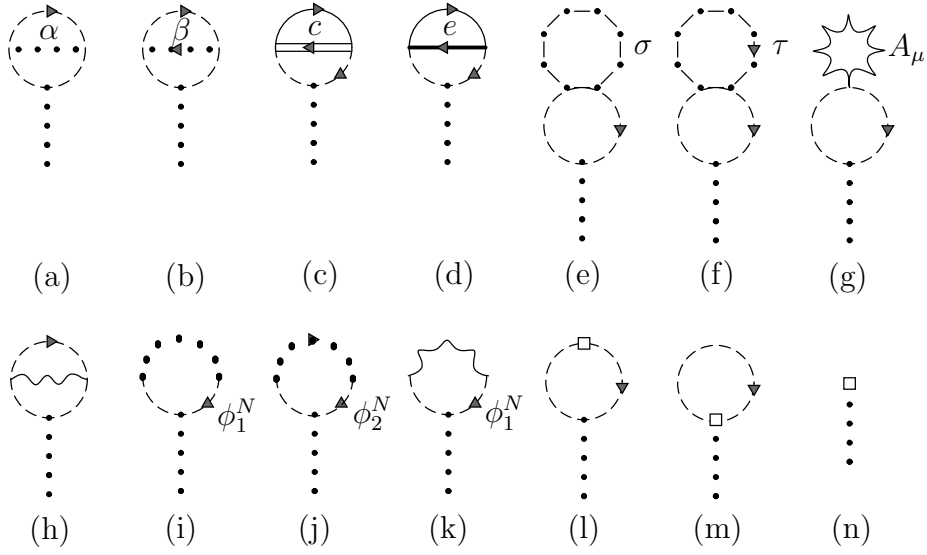}
\caption{Next-to-leading order diagrams of the $\alpha$-tadpole. The squares represent counterterm vertices.}
\label{Feynman}
\end{figure}

Next-to-leading order diagrams of the $\alpha$-tadpole ${\mit\Gamma}_{\alpha}$ are shown in Fig.~\ref{Feynman}. Figs.~2l-2n are counterterm diagrams. These diagrams receive the following contributions:
\begin{eqnarray}
&&{\mit \Gamma}_{\alpha, 2{\rm l}}=-\Delta Z_{\phi 1}^{(1)}+\Delta Z_{\phi 2}^{(1)},  \\
&&{\mit\Gamma}_{\alpha, 2{\rm m}}=\Delta (Z_{\phi 1}^{(1)}+Z_\alpha^{(1)})-\Delta (Z_{\phi 2}^{(1)}+Z_\alpha^{(1)}),\\
&&{\mit\Gamma}_{\alpha, 2{\rm n}}=Nr^2(Z_{\phi 1}^{(1)}+Z_\alpha^{(1)})-N ((Z_g^{-1})^{(1)}+Z_\alpha^{(1)})/g,
\end{eqnarray}
where
\begin{eqnarray}
\Delta=\int {d^3 p\over (2\pi)^3}{N \over p^2}.
\end{eqnarray}
The sum of Figs.~2l and 2m is zero. From the $Z$ factors (\ref{zfac2}), we find
\begin{eqnarray}
{\mit \Gamma}_{\alpha, 2{\rm n}}=-{2r^2\over \pi^2}\ln {\Lambda \over \mu}-{N\over g}Z_g^{(1)}.\label{coun}
\end{eqnarray}
Fig.~2k and the sum of Figs.~2i and 2j are
\begin{eqnarray}
{\mit \Gamma}_{\alpha, 2{\rm k}}=0,~~~{\mit \Gamma}_{\alpha, 2{\rm i}}+{\mit \Gamma}_{\alpha, 2{\rm j}}={2r^2\over \pi^2}\ln {\Lambda \over \mu}.
\end{eqnarray}
We find that this logarithmic divergence is canceled by the first term in the counterterm (\ref{coun}). Thus $Z_g^{(1)}$ can be obtained by computing Figs.~2a-2h. For Fig.~2a we obtain
\begin{eqnarray}
{\mit \Gamma}_{\alpha, 2{\rm a}}={\mit \Gamma}_{\alpha, 2{\rm a}}^{(\phi 1\,{\rm mode})}+{\mit \Gamma}_{\alpha, 2{\rm a}}^{(\phi 2\,{\rm mode})}=0,
\end{eqnarray}
because
\begin{eqnarray}
{\mit \Gamma}_{\alpha, 2{\rm a}}^{(\phi 1\,{\rm mode})}=-{\mit \Gamma}_{\alpha, 2{\rm a}}^{(\phi 2\,{\rm mode})}=\int {d^3 p\over (2\pi)^3}\int {d^3 k\over (2\pi)^3}D^\alpha(p){N \over k^4(p+k)^2}.
\end{eqnarray}
For the same reason, we have found that the each of Figs.~2b-2h is zero. Finally we obtain
\begin{eqnarray}
Z_{g}^{(1)}=0.
\end{eqnarray}
This implies that the $\beta$-function receives no contributions at next-to-leading order:
\begin{eqnarray}
\beta^{(1)}({\tilde g})=0.
\end{eqnarray}

\section{Discussion}
We have shown that the $\beta$-function in the ${\cal N}=4$ SUSY $T^*(CP^{N-1})$ model in three dimensions receives no quantum corrections to leading and next-to-leading orders. There is a theorem that the $\beta$-function in leading and next-to-leading orders has renormalization scheme independent meaning in the usual perturbation expansion \cite{Gross}. In $1/N$ expansion, however, the $\beta$-function will not probably have this feature. It is an important question whether the absence of non-leading corrections to the $\beta$-function persists to all orders in $1/N$. We need to make use of the superfield formulation in order to handle the problem. For instance, in perturbation expansion the two-dimensional ${\cal N}=4$ SUSY nonlinear sigma models were found to be finite to all orders using a general argument combining the background field method and differential geometry in the superfield formulation \cite{Alvarez}. In the $1/N$ expansion, we already know that the two-dimensional ${\cal N}=4$ SUSY  $T^*(CP^{N-1})$ model should be finite to leading order in the superfield formulation \cite{Rocek}.

\section*{Acknowledgements}
We would like to thank M. Sakamoto for a careful reading of the manuscript and enlightening discussion of the dependence of the $Z$ factors and the $\beta$-function on the scheme. This work is supported partially by the Grants in Aid of Ministry of Education, Culture and Science (Priority Area B "Supersymmetry and Unified Theory" and Basic Research C). M. Y. was supported by a Research Assistantship of Chuo University.

\end{document}